\documentclass[aps,pra,superscriptaddress,twocolumn]{revtex4-1}
\usepackage[usenames]{color}
\usepackage{pstricks}
\usepackage{amsmath}
\usepackage{amssymb}
\usepackage{subfigure}
\usepackage{afterpage}
\usepackage{nameref}
\usepackage{array}
\usepackage{multirow}
\usepackage{graphicx}
\usepackage{color}
\usepackage{placeins}

\begin{document}
\title{Adimensional theory of shielding in ultracold collisions of dipolar rotors}

\author{Maykel L.~Gonz{\'a}lez-Mart\'{\i}nez}
\affiliation{Laboratoire Aim{\'e} Cotton, CNRS,
             Universit{\'e} Paris-Sud, ENS Paris-Saclay, Universit{\'e} Paris-Saclay\\
             B{\^a}t.\ 505, Campus d'Orsay,
             91405 Orsay, France}
\author{John L. Bohn}
\affiliation{JILA, NIST, and Department of Physics, University of Colorado,
Boulder, Colorado 80309-0440, USA}
\author{Goulven Qu{\'e}m{\'e}ner}
\affiliation{Laboratoire Aim{\'e} Cotton, CNRS,
             Universit{\'e} Paris-Sud, ENS Paris-Saclay, Universit{\'e} Paris-Saclay\\
             B{\^a}t.\ 505, Campus d'Orsay,
             91405 Orsay, France}

\date{\today}
\begin{abstract}
We investigate the electric field shielding of ultracold collisions of dipolar rotors, 
initially in their first rotational excited state, using an adimensional approach.
We establish a map of good and bad candidates for efficient evaporative cooling 
based on this shielding mechanism, 
by presenting the ratio of elastic over quenching processes 
as a function of 
a rescaled rotational constant $\tilde{B} = B/s_{E_3}$ and 
a rescaled electric field $\tilde{F} = d F / B$. 
$B,d,F,s_{E_3}$ are respectively the rotational constant, the full electric dipole moment of the molecules, 
the applied electric field and a characteristic dipole-dipole energy.
We identify two groups of bi-alkali dipolar molecules. 
The first group, including RbCs, NaK, KCs, LiK, NaRb, LiRb, NaCs and LiCs,
is favorable with a ratio over 1000
at collision energies equal (or even higher) 
to their characteristic dipolar energy.
The second group, including LiNa and KRb, is not favorable.
More generally, for molecules
well described by Hund's case b,
our adimensional study provides the conditions of
efficient evaporative cooling. The range of appropriate rescaled 
rotational constant and rescaled field is approximately $\tilde{B} \ge 10^8$
and $3.25 \le \tilde{F} \le 3.8$,
with a maximum ratio reached for $\tilde{F} \simeq 3.4$ for a given $\tilde{B}$.
We also discuss the importance of the electronic van der Waals interaction
on the adimensional character of our study.
\end{abstract}


\maketitle
\section{Introduction}
\label{sec:introduction}

Ultracold dipolar molecules have been the subject of
tremendous experimental and theoretical investigations
these past years.
They are promising candidates for many interesting 
applications \cite{Carr_NJP_11_055049_2009},
from many-body physics \cite{Baranov_CR_112_5012_2012}
to ultracold controlled chemistry 
\cite{Gonzalez-Martinez_PRA_90_052716_2014,Docaj_PRL_116_135301_2016},
from quantum information \cite{Yelin_BookChapter_2009} 
to precision measurements \cite{Tarbutt_BookChapter_2009}.
Different kinds of ultracold dipolar molecules exist now.
They have been produced in ultracold gases with high enough 
densities to study their two-body interactions 
and collisions \cite{Quemener_CR_112_4949_2012}.
They can possess an electric or magnetic dipole moment 
so that they can be controlled by either an electric or magnetic field.
These molecules can be fermionic or bosonic, chemically reactive 
or not, and produced in the absolute ground state 
or in a weakly bound state. 
Examples of fermionic molecules are 
$^{40}$K$^{87}$Rb \cite{Ni_S_322_231_2008} and 
$^{23}$Na$^{40}$K \cite{Park_PRL_114_205302_2015} 
while 
$^{87}$Rb$^{133}$Cs \cite{Takekoshi_PRL_113_205301_2014,Molony_PRL_113_255301_2014} and 
$^{23}$Na$^{87}$Rb \cite{Guo_PRL_116_205303_2016} are examples of bosonic molecules.
They were produced in their absolute ground state
in which they possess an electric dipole moment
and can then be controlled by an electric field \cite{Ni_N_464_1324_2010,DeMiranda_NP_7_502_2011}.
Magnetic dipolar ultracold 
molecules such as Er$_2$ have also been produced 
in a weakly bound state. They possess a magnetic dipole moment 
and can be controlled with a magnetic field \cite{Frisch_PRL_115_203201_2015}.

However, all these molecules share the same problem: 
they suffer from two-body collisional losses (quenching)
whether it is due to the chemical reactivity of the molecules
\cite{Ospelkaus_S_327_853_2010,Zuchowski_PRA_81_060703_2010}, 
inelastic collisions to lower molecular states \cite{Frisch_PRL_115_203201_2015}, or
possible collisional losses mediated by long-lived complex
\cite{Mayle_PRA_85_062712_2012,Mayle_PRA_87_012709_2013}.
It is then problematic to reach the quantum degeneracy of an ultracold gas of
dipolar molecules. Quantum degeneracy can be reached by evaporative cooling,
which was succesfully applied to obtain Bose--Einstein condensates
of ultracold neutral atoms \cite{Cornell_RMP_74_875_2002, Ketterle_RMP_74_1131_2002}
and degenerate Fermi gases \cite{DeMarco_S_285_1703_1999}.
The technique relies (at least but not only) on large two-body elastic rate coefficients
for fast thermalization times and on small quenching rate coefficients 
for low collisional losses.
Therefore, shielding the molecules from these unwanted collisional losses is
absolutely essential to reach quantum degeneracy in ultracold gases.

A somewhat counter-intuitive scheme has been proposed to shield polar molecules from quenching collisions, by preparing them in their first rotationally excited state.  In this state, if the electric field is tuned just above a critical value, there results an effective repulsion that keeps the molecules from changing their internal state
or reacting.
This has been studied for inelastic collisions
\cite{Avdeenkov_PRA_73_022707_2006},
reactive collisions of $^1 \Sigma$ molecules \cite{Wang_NJP_17_035015_2015} 
or $^2 \Sigma$ molecules \cite{Quemener_PRA_93_012704_2016}.
In contrast with these previous works, 
this paper presents a systematic study
using an adimensional perpective. We determine adimensional rescaled parameters
that govern the dynamics of the systems, 
namely a rescaled rotational constant, 
a rescaled electric field and a rescaled collision energy.
Then, all molecules are treated on equal footing 
with the same rescaled formalism
\cite{Bohn_PRA_63_052714_2001,Bohn_NJP_11_055039_2009,Ticknor_PRA_80_052702_2009,
Gao_PRL_105_263203_2010,Wang_PRA_85_022704_2012}.
We find the molecules and the range of the rescaled parameters 
for which the collisional loss suppression is low enough so that 
evaporative cooling techniques can be used efficiently 
to reach quantum degeneracy in ultracold gases of dipolar molecules.

The paper is organized as follows. 
In Sec.~\ref{sec:theory} we briefly recall the formalism used in the former papers
\cite{Avdeenkov_PRA_73_022707_2006,Wang_NJP_17_035015_2015,Quemener_PRA_93_012704_2016}
using dimensional quantities. 
Then we introduce the adimensional formalism 
based on the dipolar interaction which defines a characteristic length and energy. 
We obtain adimensional rescaled cross sections, rate coefficients and scattering length
as a function of the rescaled parameters.
We present and discuss our results in Sec.~\ref{sec:results+dicussion}. 
With a single figure, one can determine
the good molecular candidates for efficient evaporative cooling
based on the shielding. 
We also discuss the importance of the electronic van der Waals interaction
on the adimensional character of our study.
Finally, we conclude in Sec.~\ref{sec:conclusion}.

\section{Theory}
\label{sec:theory}

\subsection{Presentation of the scattering problem using dimensional quantities}
\label{sec:subsecA}

We consider collisions between two species 1 and 2 of mass $m_1$ and $m_2$,
in the presence of an external electric field $F$. 
The direction of the electric field is chosen
as the space-fixed quantization axis. 
The species 1 and 2 are diatomic molecules
considered in this study as dipolar rotors
with permanent electric dipole moment $d_1 = d_2 = d$.
We do not consider the hyperfine structure.
The scattering Hamiltonian can be written
\begin{eqnarray}
 \hat{\mathcal{H}} = - \frac{\hbar^2}{2\mu} r^{-1} \frac{d^2}{dr^2} r
                     + \frac{\hat{l}^2}{2\mu r^2}
                     + \hat{\mathcal{H}}_1 + \hat{\mathcal{H}}_2
                     + \hat{\mathcal{V}},
 \label{eq:Heff}
\end{eqnarray}
where $\mu = m_1 m_2 / (m_1+m_2)$ is the reduced mass for the molecule-molecule collision, 
$r$ is the distance between
the species' centers of mass, and $\hat{l}$ is the space-fixed operator for the
orbital angular momentum between the two species. 
$\hat{\mathcal{H}}_1$ and $\hat{\mathcal{H}}_2$ describe the 
Hamiltonian of the isolated species 1 and 2, 
including their interactions with the applied field.
$\hat{\mathcal{V}} = \hat{\mathcal{V}}_\mathrm{el} + \hat{\mathcal{V}}_\mathrm{dip}$ 
contains all interactions between the species, with
contributions that include the electronic potential $\hat{\mathcal{V}}_\mathrm{el}$ 
and the dipole-dipole interaction $\hat{\mathcal{V}}_\mathrm{dip}$.
In Sec. \ref{sec:subsecB}, Sec. \ref{sec:subsecC}, and Sec. \ref{sec:subsecD}, 
we will reduce the formalism to a model
that consists in taking the long-range interaction of the molecules
and treating the short-range interaction using an absorbing potential. 
The Hamiltonian for an isolated dipolar rotor is
\begin{equation}
 \hat{\mathcal{H}}_\mathrm{1,2} \equiv \hat{\mathcal{H}}_\mathrm{mol} = 
   \frac{\hat{n}^2}{2 I}  - \hat{d} \cdot \hat{{F}}
  = \frac{\hat{n}^2}{2 I} - d \,{F}  \cos\theta ,
\end{equation}
where $\hat{n}$ is the rotational angular momentum and 
$I$ the moment of inertia of the dipolar rotor.
The corresponding rotational constant is related to the moment of inertia by $B = \hbar^2/2I$.
The angle $\theta$ corresponds to the angle between the permanent dipole moment 
and the electric field.
The dipolar interaction is 
\begin{equation}
 \hat{\mathcal{V}}_\mathrm{dip} = -\frac{\sqrt{6}}{4\pi\epsilon_0} r^{-3} \
  \mathrm{T}^2(\hat{d_1},\hat{d_2}) \cdot \mathrm{T}^2(\boldsymbol{u}_r) ,
\end{equation}
where $\mathrm{T}^k$ represents a spherical tensor of rank $k$ and
$\boldsymbol{u}_r$ is a unit vector in the direction of $\boldsymbol{r}$.
We solve the quantum-mechanical scattering problem
using the coupled-channel method. The total wave function is first expanded in
a set of $N$ conveniently chosen basis functions $|i \rangle$,
\begin{eqnarray}
 |\Psi(r,\xi)\rangle = r^{-1} \sum_{i} \chi_{i}(r) \, |i\rangle ,
 \label{eq:Psi}
\end{eqnarray}
where $\xi$ is a collective variable including all coordinates except $r$, and
$i$ is the set of quantum numbers that label the basis functions.  Each
different combination of quantum numbers $i$ defines a channel. 
We choose for the individual species the set of bare basis functions 
$|\alpha_{1,2}\rangle \equiv |\alpha\rangle = |n \, m_n\rangle$, so that
\begin{equation}
 \langle \alpha| \hat{\mathcal{H}}_\mathrm{mol} |\alpha' \rangle 
  = B \, n(n+1) \, \delta_{\alpha, \alpha'}
 - d \, {F} \langle \alpha| \cos\theta |\alpha' \rangle.
 \label{eq:Hrot}
\end{equation}
The eigenfunctions of the corresponding matrix 
become the dressed internal states
$|\tilde{\alpha}_{1,2}\rangle \equiv |\tilde{\alpha}\rangle = |\tilde{n} \, m_n\rangle $. 
The projection quantum number $m_n$ remains a good quantum number
while $n$ is not. $\tilde{n}$ indicates that 
the dressed state $|\tilde{n} \, m_n\rangle $ has a main 
character in $n$ when $F \simeq 0$, 
but is in general a linear combination of the bare states $|n \, m_n\rangle$.
The corresponding eigenvalues of the dipolar rotors 1 and 2 
in the field are $E_{\tilde{\alpha}_1}$ and $E_{\tilde{\alpha}_2}$. 
The basis functions are then symmetrized in terms of $|\tilde{\alpha}_{1,2}\rangle$,
adding the orbital angular momentum, so that $|i\rangle $ is defined in Eq.~\eqref{eq:Psi} by
\begin{equation}
 |i\rangle \equiv
  \frac{1}{\sqrt{2(1 + \delta_{\tilde{\alpha}_1 \tilde{\alpha}_2})}}
  \left(|\tilde{\alpha}_1 \rangle |\tilde{\alpha}_2\rangle 
  + \eta |\tilde{\alpha}_2 \rangle |\tilde{\alpha}_1\rangle\right)|l m_l\rangle.
  \label{eq:kket}
\end{equation}
The corresponding energy of the channel $|i\rangle$ 
is $E_{i} = E_{\tilde{\alpha}_1} + E_{\tilde{\alpha}_2}$, the energy of the 
two separated dipolar rotors 1 and 2 in the field.
$\eta = + 1$ ($\eta = - 1$) corresponds to a symmetric (anti-symmetric) function
with respect to permutation $\hat{P}$ of the identical species 1 and 2. 
The permutation operator acts on the basis function 
as $\hat{P} |i\rangle = \eta \, (-1)^l |i\rangle$.
On the other hand, 
from the symmetrization principle, $\hat{P} |\Psi\rangle = \epsilon_P |\Psi\rangle$,
where $\epsilon_P = +1$ for identical bosons and $\epsilon_P = -1$ for identical fermions.
This condition implies from Eq.~\eqref{eq:Psi} that $\hat{P} |i\rangle = \epsilon_P |i\rangle$
and imposes the selection rule $ \eta \, (-1)^l = \epsilon_P $.
The time-independent Schr{\"o}dinger equation 
$\hat{\mathcal{H}} | \Psi \rangle = E_\mathrm{tot}  | \Psi \rangle $,
for the scattering wave function $|\Psi\rangle$
expanded over the dummy argument $i'$ and 
for a total energy $E_\mathrm{tot}$ of the colliding system, 
provides a set of N by N coupled differential equations 
for the channel functions $\chi_{i'}(r)$
when projected onto the N possible bra $\langle i|$,
\begin{eqnarray}
\langle i | ( \hat{\mathcal{H}} - E_\mathrm{tot} ) | \Psi \rangle &=& 
\sum_{i'} \langle i | ( \hat{\mathcal{H}} - E_\mathrm{tot} ) | r^{-1}  \chi_{i'}(r) |i'\rangle \nonumber \\ 
&=& 0. 
\end{eqnarray}
Using the form of the Hamiltonian in Eq.~\eqref{eq:Heff}, we get the following set
of coupled equations,
\begin{multline}
 \left[ - \frac{\hbar^2}{2\mu} \frac{d^2}{d{r}^2}
        + \frac{\hbar^2 l(l + 1)}{2\mu {r}^2}
        + E_{i} - E_\mathrm{tot}  \right] \chi_i({r}) \\
         + \sum_{i'} V_{i,i'}(r) \, \chi_{i'}({r})  = 0 ,
 \label{eq:coupleddim}
\end{multline}
where we used the notation $V_{i,i'} \equiv \langle i| \hat{\mathcal{V}} |i'\rangle$.
The total energy $E_\mathrm{tot} = {E}_\mathrm{init} + E_c$ is the sum of the initial 
combined molecular state and the collision energy $E_c$.
Each term of the equations has dimensions of energy.
We will now get rid of the dimensional character of the equations.

\subsection{The adimensional scattering problem}
\label{sec:subsecB}

\begin{figure}[t]
 \includegraphics[width=86mm]{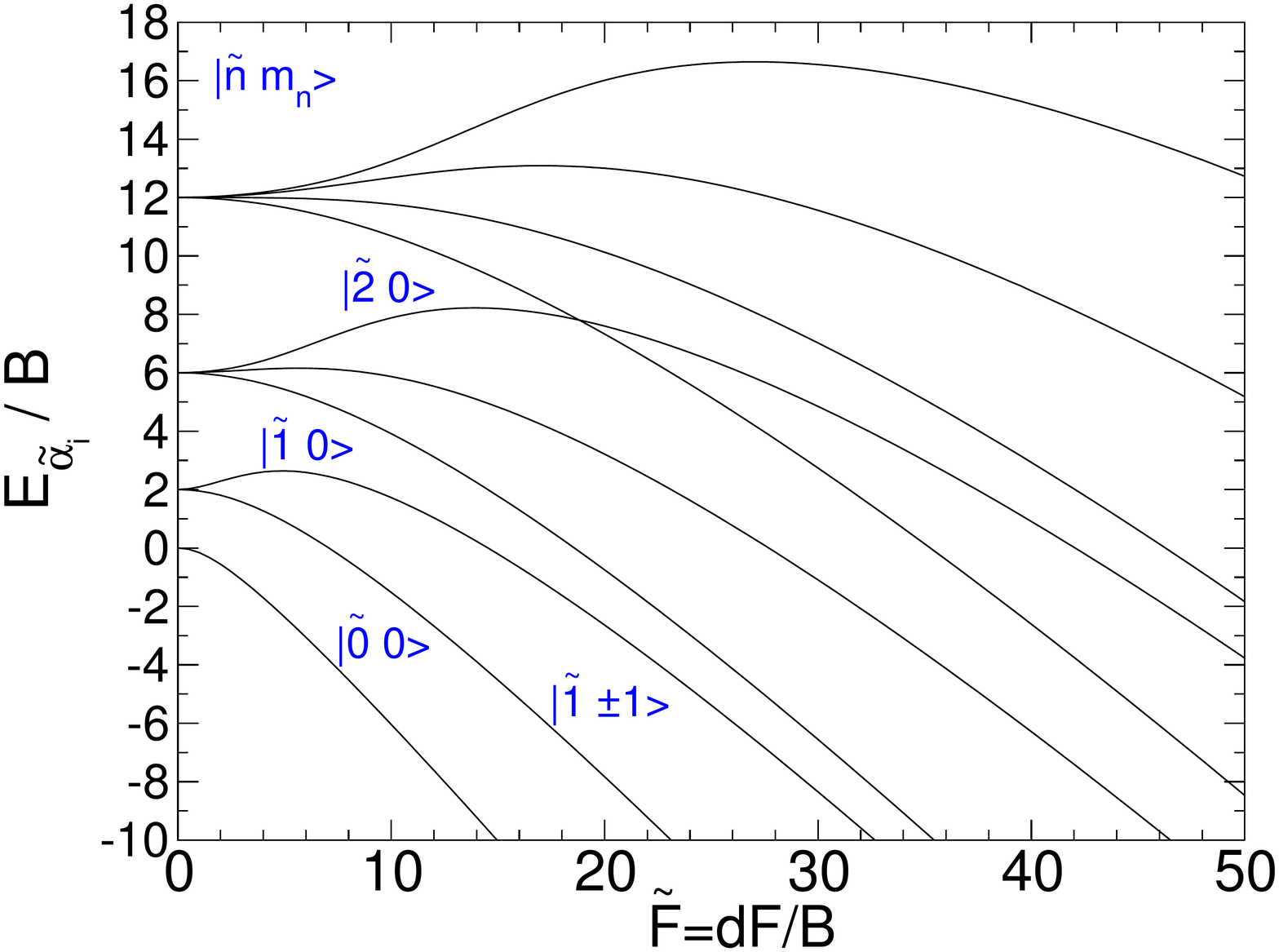}
 \caption{(Color online). Rescaled energies ${E}_{\tilde{\alpha_i}}/B$ of a dipolar rotor as a function of the rescaled field $\tilde{F} = dF/B$. Some dressed states 
$|\tilde{n}_i m_{n_i} \rangle$ are explicitely indicated.}
 \label{fig:nrg1}
\end{figure}
\begin{figure}[t]
 \includegraphics[width=86mm]{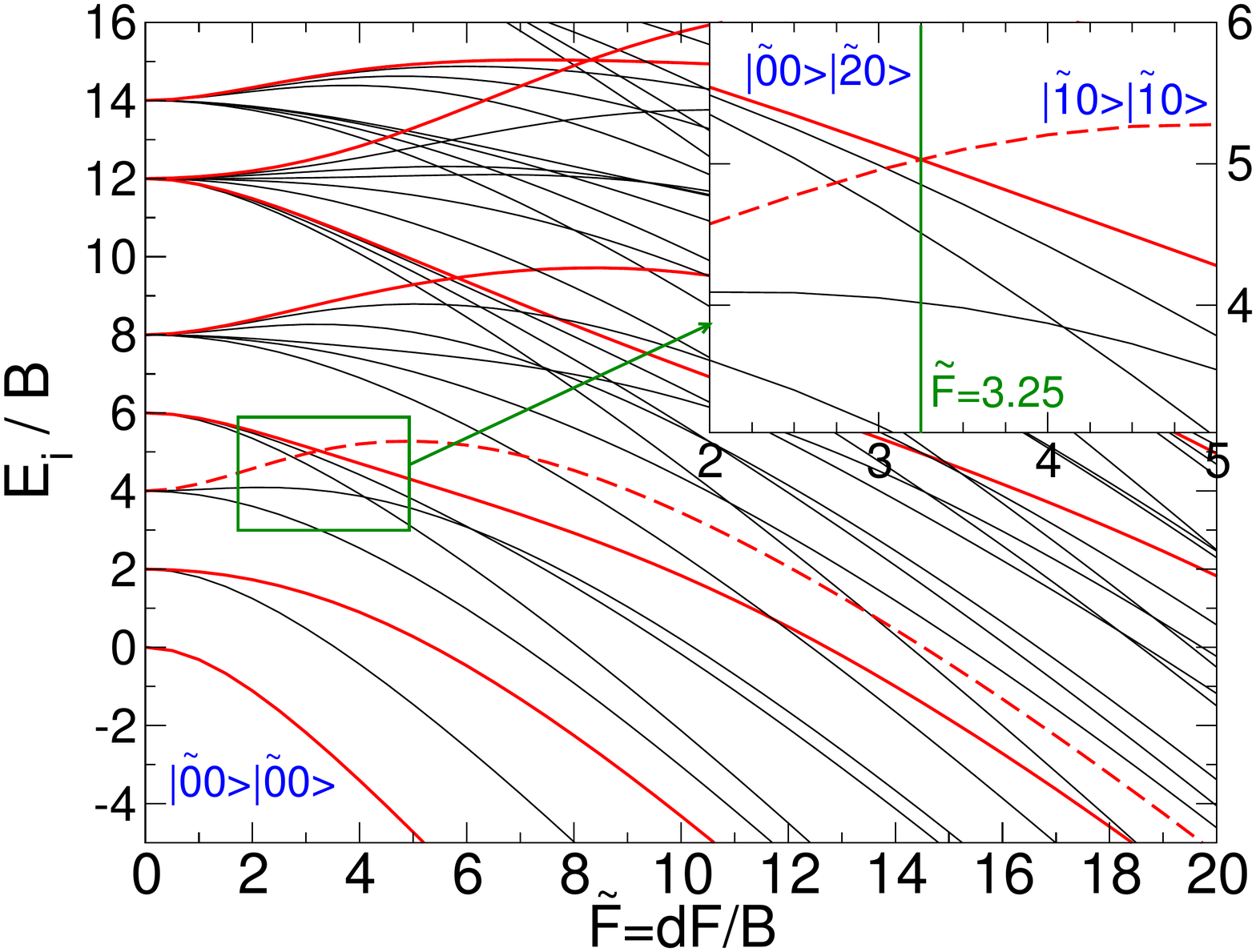}
 \caption{(Color online). Rescaled energies ${E}_i/B$ of two combined dipolar rotors 
 as a function of the rescaled field $\tilde{F} = dF/B$. Some combined dressed states 
$|\tilde{n}_1 m_{n_1} \rangle| \tilde{n}_2 m_{n_2} \rangle$ are explicitely indicated.
The field at which the initial state $|\tilde{1}0\rangle|\tilde{1}0\rangle$ (red bold dashed line) crosses the $|\tilde{0}0\rangle|\tilde{2}0\rangle$ one is $\tilde{F}=3.25$.
The red bold dashed and plain lines correspond to states with $m_{n_1}=m_{n_2}=0$.}
 \label{fig:nrg2}
\end{figure}

The adimensional problem is set up by defining a typical characteristic length and energy
from the form of the interaction.
A $C_n/r^{n}$ type interaction defines characteristic length and energy scales 
\cite{Gao_PRL_105_263203_2010}
\begin{equation}
 s_{r_n} \equiv \left(\frac{2 \mu C_n}{\hbar^2} \right)^{\frac{1}{n-2}},\;
 s_{E_n} \equiv \frac{\hbar^2}{2 \mu \, s^2_{r_n}}.
\end{equation}
As the physics of the shielding
occurs due to the dipole-dipole interactions 
at nonzero applied electric field at a large $r$, 
the dominant and most relevant energy scale of the theory is the dipolar interaction.
The matrix representation of this interaction 
in $\langle i| \hat{\mathcal{V}} |i'\rangle$ can be written as
\begin{equation}
 \langle i| \hat{\mathcal{V}}_\mathrm{dip} |i'\rangle 
  = \frac{C_3}{r^3} \, \zeta_{i,i'}(l, l', m_l, m_l';F) ,
\end{equation}
with $C_3 \equiv \frac{d^2}{4\pi\epsilon_0}$, and $\zeta_{i,i'}(l, l', m_l, m_l';F)$ 
being adimensional geometrical coefficients depending on the orbital angular momentum
as well as the rotational angular momentum
via the dressed states $i,i'$.  
We also indicated the implicit dependence of the coefficients on $F$.
Hence, the length scale for the $r^{-3}$ dipolar interaction term is
\begin{eqnarray}
 s_{r_3} &\equiv& \frac{2 \mu C_3}{\hbar^2}  
 = \frac{2\mu}{\hbar^2} \frac{d^2}{4\pi\epsilon_0} \nonumber \\
 &=& 2(\mu/\mathrm{a.u.}) (d/\mathrm{a.u.})^2 ,
\end{eqnarray}
and the corresponding energy scale is
\begin{eqnarray}
 s_{E_3} &\equiv& \frac{\hbar^2}{2 \mu \, s^2_{r_3}} = \frac{\hbar^6}{(2 \mu)^3 \, (d^2/4\pi\epsilon_0)^2} \nonumber \\
 &=& \left[ 8(\mu/\mathrm{a.u.})^3 (d/\mathrm{a.u.})^4 \right]^{-1}.
\end{eqnarray}
This defines adimensional lengths, $\tilde{r} \equiv r/s_{r_3}$ and energies
$\tilde{E} \equiv E/s_{E_3}$.
If one includes only $\hat{\mathcal{V}}_\mathrm{dip}$ in $\hat{\mathcal{V}}$, 
the rescaled coupled-channel equation for a given channel $|i\rangle$ becomes
\begin{multline}
 \left[ - \frac{d^2}{d\tilde{r}^2}
        + \frac{l(l + 1)}{\tilde{r}^2}
        + \left(\frac{{E}_{i} - {E}_{\mathrm{init}}}{B}\right) \, \tilde{B} - \tilde{E}_c \right] \chi_i(\tilde{r})  \\ 
        + \sum_{i'} \left(\frac{\zeta_{i,i'}(l, l', m_l, m_l';\tilde{F})}{\tilde{r}^3} \right) \chi_{i'}(\tilde{r}) 
           = 0 .
 \label{eq:coupledadim}
\end{multline}
There are as many equations as channels $|i\rangle$. 
The above adimensional equations display three adimensional quantities relating 
to the physical parameters of the colliding system.
The first one,
\begin{eqnarray}
 \tilde{B} &=& \frac{B}{s_{E_3}}  = \frac{8 B \mu^3}{\hbar^6} 
 \left(\frac{d^2}{4\pi\epsilon_0}\right)^2  \nonumber \\
 &=& 8 (B/\mathrm{a.u.}) (\mu/\mathrm{a.u.})^3 (d/\mathrm{a.u.})^4 ,
 \label{eq:Btilde}
\end{eqnarray}
represents the rotational constant rescaled 
over the dipolar energy. It depends on the rotational constant, the reduced mass and 
the full electric dipole moment of the system.
Therefore the first parameter contains all the information of an individual dipolar rotor
and is fixed for a given system.
The second parameter in Eqs.~\eqref{eq:coupledadim} is
\begin{eqnarray}
  \frac{{E}_{i} - {E}_{\mathrm{init}}}{B} .
 \label{eq:Ektilde}
\end{eqnarray}
It represents the difference between the energy ${E}_{i}$
of two separated dipolar rotors in channel $|i\rangle$ 
and the energy ${E}_{\mathrm{init}}$ of the initial state, 
rescaled by the rotational constant. 
The individual rescaled energies 
${E}_{\tilde{\alpha}_{1,2}}/B$ are obtained directly from 
Eq.~\eqref{eq:Hrot} when the equation is divided on both sides 
by the rotational constant $B$ \cite{Meyenn_ZP_231_154_1970}.
They are function of a rescaled field defined by
\begin{eqnarray}
\tilde{F} = \frac{d \, F}{B} ,
 \label{eq:Ftilde}
\end{eqnarray}
and plotted in Fig.~\ref{fig:nrg1} as a function of $\tilde{F}$.
The rescaled energies ${E}_i/B$ are plotted in Fig.~\ref{fig:nrg2} as a function 
of $\tilde{F}$.
In the following, as the rescaled energies, their differences, 
as well as the coefficients $\zeta_{i,i'}$
are fixed for a given rescaled field, 
the second parameter is monitored implicitely by the 
rescaled field $\tilde{F}$ in which the system is colliding.
The third parameter in Eqs.~\eqref{eq:coupledadim},
\begin{eqnarray}
\tilde{E}_c = \frac{E_c}{s_{E_3}} ,
 \label{eq:Etottilde}
\end{eqnarray} 
corresponds to the collision energy rescaled over the dipolar energy.
In this study, we will consider the ultracold regime so that 
$\tilde{E}_c \to 0$ and is fixed for a given initial state of the system.
The adimensional equations depend solely on $\tilde{B}$ 
and $\tilde{F}$.

\subsection{Adding the electronic van der Waals interaction}
\label{sec:subsecC}

Eqs.~\eqref{eq:coupledadim} are useful to determine which parameters are the relevant ones
using an adimensional perspective. However, this is possible because we have only used 
the dipolar interaction as the typical interaction, which defined the proper length and energy scales $s_{r_3}$ and $s_{E_3}$. In practice we also have to include the electronic interaction.
We use a simple long-range, isotropic description based on the leading dispersion term.
The matrix representation of the electronic interaction 
in $\langle i| \hat{\mathcal{V}} |i'\rangle$ is then
\begin{eqnarray}
 \langle i| \hat{\mathcal{V}}_\mathrm{el} |i'\rangle \approx \delta_{i,i'}\frac{C_6^{\mathrm{el}}}{r^6} ,
\end{eqnarray}
where $C_6^{\mathrm{el}}$ represents the electronic 
van der Waals coefficient between the two dipolar rotors. 
It must be noted, however, that this term is not rigorously scale-free in $\tilde{r}$
for it defines a different characteristic length
\begin{eqnarray}
 s_{r_6} \equiv \left(\frac{2\mu}{\hbar^2} |C_6^{\mathrm{el}}| \right)^{\frac{1}{4}} \neq s_{r_3}.
\end{eqnarray}
Therefore, we cannot end up in general with a strictly adimensional study.
Neglecting the electronic van der Waals term is also not possible 
given that in some cases the geometrical factor $\zeta_{i,i'}$ 
vanishes in the diagonal element of 
$\langle i| \hat{\mathcal{V}}_\mathrm{dip} |i'\rangle$, 
such as for an incoming and outgoing $s$-wave $l=l'=0$.
However, if the electronic van der Waals term plays a negligible role
in the shielding effect, the study can be considered adimensional.
In practice, we don't use the adimensional coupled 
equations Eqs.~\eqref{eq:coupledadim}
to compute the scatering properties.
We instead solve the dimensional coupled equations 
in Eqs.~\eqref{eq:coupleddim}
for a fixed electronic $C_6^{\mathrm{el}}$ coefficient and 
appropriately come back to adimensional rescaled quantities. 
At the end of Sec.~\ref{sec:results+dicussion}, 
we discuss in more detail for which systems
the electronic van der Waals interactions play a negligible part
and therefore when the study becomes adimensional.

\subsection{Cross sections, rate coefficients and scattering length}
\label{sec:subsecD}

The close-coupling equations are solved for each $r$ from a minimum value 
$r_\mathrm{min}$ to a maximum value $r_\mathrm{max}$ using a log-derivative propagation method
\cite{Johnson_JCP_13_445_1973,Manolopoulos_JCP_85_6425_1986}.
At $r_\mathrm{min}$, we initialize the propagation by a complex, diagonal log-derivative matrix 
$\mathbf{Z}$ whose elements are given by
\cite{Wang_NJP_17_035015_2015}
\begin{multline}
 Z(r = r_\mathrm{min}) = \\
    \frac{{k}_\mathrm{min} (4sc\sqrt{1-p_\mathrm{SR}} - i \, p_\mathrm{SR})}
        {c^2\left(\sqrt{1-p_\mathrm{SR}} - 1\right)^2
         + s^2\left(\sqrt{1-p_\mathrm{SR}} + 1\right)^2},
 \label{eq:logder}
\end{multline}
where
\begin{eqnarray}
 {k}_\mathrm{min} = \sqrt{\frac{2 \mu}{\hbar^2}\bigg[E_\mathrm{tot} - \bigg(V_{i,i}(r_\mathrm{min})+\frac{\hbar^2 l(l + 1)}{2\mu {r_\mathrm{min}}^2}\bigg)\bigg]},
\end{eqnarray}
$c = \cos{({k}_\mathrm{min} r_\mathrm{min} + \delta_\mathrm{SR})}$ and
$s = \sin{({k}_\mathrm{min} r_\mathrm{min} + \delta_\mathrm{SR})}$.
$0 \le p_\mathrm{SR} \le 1$ and 
$0 \le \delta_\mathrm{SR} \le \pi$ are two parameters
that tune the loss probability 
and the phase shift 
of the incoming flux at $r_\mathrm{min}$.
This is as if we had approximated 
the ``chemically-active'' internal 
configuration region ($r \le r_\mathrm{min}$) 
of each channel $|i\rangle$ 
by a square-well potential from $r=0$ to $r=r_\mathrm{min}$, 
whose depth is given by
\begin{eqnarray}
\frac{C_3}{r_\mathrm{min}^3} \, \zeta_{i,i} + \frac{C_6^{\mathrm{el}}}{r_\mathrm{min}^6} 
+\frac{\hbar^2 l(l + 1)}{2\mu {r_\mathrm{min}^2}},
\end{eqnarray}
and whose corresponding log-derivative
is Eq.~\eqref{eq:logder} at $r = r_\mathrm{min}$.
At the end of the propagation, one usually obtains the scattering matrix $\mathbf{S}$ 
by applying asymptotic boundary conditions at $r_\mathrm{max}$ when $V_{i,i}(r = r_\mathrm{max}) \to 0$.
As we start with an arbitrary complex log-derivative in Eq.~\eqref{eq:logder} mimicking 
a phenomenological loss at short-range, it implies that the $\mathbf{S}$ matrix
is not necessarily unitarity.
The diagonal element of a given column  
determines the magnitude of the elastic process
while the sum of the off-diagonal terms determines the inelastic processes. 
The (positive) difference of unity with the sum of the modulus square 
of the elements of a matrix column 
determines the (phenomenological) loss processes. In our study, we do not distinguish
between loss and inelastic processes, they all contribute to destruction or removal of 
the molecules in an experimental trap. Therefore, we consider only the quenching processes
which are the sum of the inelastic and the loss processes.

As we are interested in scattering properties that are independent 
of the collision energy, it is then more useful to present and
compute the scattering length
instead of the cross sections or the rate coefficients. 
The $s$-wave scattering length 
becomes a constant when the wavevector ${k}  = \sqrt{2 \mu E_c / \hbar^2} \to 0$.
It is defined as \cite{Hutson_NJP_9_152_2007}
\begin{eqnarray}
a &=& a_\mathrm{re} - i \, a_\mathrm{im} \nonumber \\ 
  &=& \frac{1}{i \, {k}} 
  \left( \frac{1-S_{00}({k})}{1+S_{00}({k})} \right) \bigg|_{{k} \to 0} ,
\end{eqnarray}
with $a_\mathrm{im} \ge 0$.
$S_{00}$ represents the diagonal elements of the $\mathbf{S}$ matrix corresponding to the 
initial collisional state taken into consideration.
The cross sections and rate coefficients are related to the scattering length by
\begin{align}
\sigma_\mathrm{el} &= 4 \pi |a|^2 \times \Delta &
\sigma_\mathrm{qu} &= \frac{4 \pi \, a_\mathrm{im}}{{k}} \times \Delta  \\
\beta_\mathrm{el} &= \frac{4 \pi \hbar {k} |a|^2}{\mu} \times \Delta &
\beta_\mathrm{qu} &= \frac{4 \pi \hbar \, a_\mathrm{im}}{\mu} \times \Delta ,
\end{align}
where $\Delta=2$ if the particles are identical and start in indistinguishable states,
$\Delta=1$ otherwise.
Since we are using the dimensional equations Eqs.~\eqref{eq:coupleddim} 
we have to rescale the quantities so that they are adimensional. 
This is done by dividing the scattering length with the characteristic length $s_{r_3}$, 
$\tilde{a} = \tilde{a}_\mathrm{re} - i \, \tilde{a}_\mathrm{im} = a/s_{r_3}$.
Similarly, we get the rescaled cross sections 
$\tilde{\sigma} = {\sigma}/s_{\sigma_{3}}$
using a characteristic cross section $s_{\sigma_{3}} = 4 \pi s_{r_3}^2$.
The rescaled rate coefficients 
$\tilde{\beta} = \beta/s_{\beta_3}$ are obtained by
using a characteristic rate coefficient 
$s_{\beta_3} = s_{\sigma_{3}} \times s_{v_{3}}$,
where $s_{v_{3}} = \hbar/\mu s_{r_3}$ corresponds to a characteristic velocity. 
The rescaled scattering length, cross sections and rate coefficients are now related by
\begin{align}
\tilde{\sigma}_\mathrm{el} &= |\tilde{a}|^2 \times \Delta &
\tilde{\sigma}_\mathrm{qu} &= \frac{\tilde{a}_\mathrm{im}}{{{\tilde{k}}}} \times \Delta \\
\tilde{\beta}_\mathrm{el} &= {\tilde{k}} \, |\tilde{a}|^2 \times \Delta & 
\tilde{\beta}_\mathrm{qu} &= \tilde{a}_\mathrm{im} \times \Delta ,
\label{eq:rescaledscat}
\end{align}
where ${\tilde{k}} = \sqrt{\tilde{E}_c} = \sqrt{E_c/s_{E_3}} \to 0$.
Note that $\tilde{k}$ is characterized via $s_{E_3}$ by the full dipole moment
of the molecule, measured in a body-fixed frame (molecular frame).
In reality, what one observes in a space-fixed frame (laboratory frame)
is the expectation value of the dipole moment, namely the 
induced dipole moment $d_\mathrm{ind}$, for a given applied electric field.
Therefore, a more apropriate characteristic 
energy that quantifies the dipolar interaction 
between the molecules is the one using the induced dipole moment
instead of the full dipole moment.
This energy depends on the applied electric field $\tilde{F}$
and characterizes the typical energy
at and below which the quantum regime is reached, 
typically when indistinguishable bosons collide 
in the single partial wave $l=0$ ($s$-wave) 
or indistinguishable fermions in the single 
partial wave $l=1$ ($p$-wave).
We note this ``quantum regime'' energy $E_\mathrm{QR}(\tilde{F})$,
and the limit of validity ${\tilde{k}} \to 0$ above corresponds to the condition 
$E_c \le E_\mathrm{QR}(\tilde{F})$.

Finally, an important quantity for experiments is the ratio $\gamma$ 
of the elastic over the quenching cross section or rate coefficient. This ratio is given by
\begin{eqnarray}
\gamma = \frac{{\beta}_\mathrm{el}}{{\beta}_\mathrm{qu}} = \frac{{\sigma}_\mathrm{el}}{{\sigma}_\mathrm{qu}} 
= \frac{|{a}|^2}{{a}_\mathrm{im}} \, {{k}}
= \frac{|\tilde{a}|^2}{\tilde{a}_\mathrm{im}} \, {\tilde{k}}.
\label{eq:ratio}
\end{eqnarray}
This ratio determines the efficiency
of the evaporative cooling technique in order to reach the quantum degeneracy
of ultracold gases.

\begin{figure*}[t]
 \includegraphics[width=200mm]{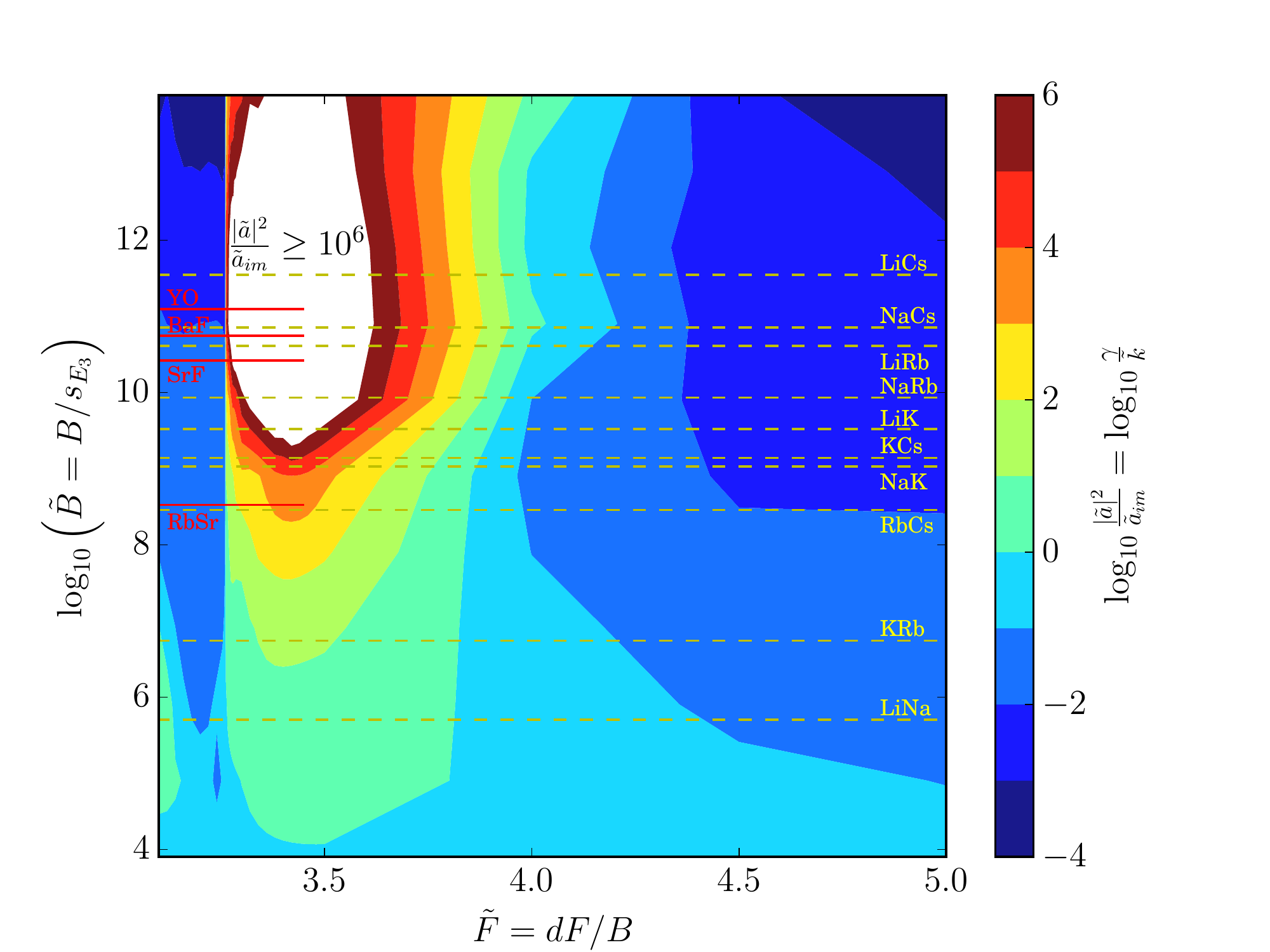}
 \caption{(Color online). $|\tilde{a}|^2/\tilde{a}_\mathrm{im} \equiv \gamma/\tilde{k}$ 
 as a function of 
 $\tilde{B}$ and $\tilde{F}$. The color scale, presented at the right of the picture, goes from $10^{-4}$ to $10^{6}$. The white area corresponds to values $\ge 10^{6}$. The $\tilde{B}$ values of some characteristic dipolar molecules are also included.}
 \label{fig:asqoveraim}
\end{figure*}

\section{Results and discussion}
\label{sec:results+dicussion}

\begin{table*}[t]
\setlength{\extrarowheight}{4pt}
\begin{center}
\caption{Summary of the different systems' parameters obtained 
from the reduced mass $\mu$, the rotational
constant $B$ \cite{Zuchowski_PRA_87_022706_2013}, 
and the full electric dipole moment $d$ \cite{Byrd_PRA_86_032711_2012} 
for bosonic $^1\Sigma$ molecules and bosonic $^2\Sigma$ molecules 
(see references inside Ref. \cite{Quemener_PRA_93_012704_2016}).
$s_{r_3}, s_{E_3}, s_{\sigma_3}, s_{\beta_3}$ are respectively
the characteristic length, energy, cross section and rate 
coefficient for the dipolar interaction  (see text for definitions).
$E_{QR} \simeq 3500 \, s_{E_3}$ is the characteristic quantum regime energy at
the field $\tilde{F}=3.4$ where the $s$-wave is predominant. 
$\tilde{B}$ is the rescaled rotational constant.
$F_{\tilde{F}=[3.25 - 3.8]}$ is the approximate range of the electric field 
where the ratio $\gamma$ would be favorable
for successful evaporative cooling. 
The systems are ordered in increasing values of $\tilde{B}$.
We provide useful conversion factors:
1 a.u. $\simeq$ 1822.88 a.m.u.; 1 a.u. $\simeq$ 219475 cm$^{-1}$; 
1 a.u. $\simeq$ 2.5417 D; 1 $a_0$ $\simeq$ 0.529 10$^{-10}$ m; 
1 a.u. $\simeq$ 315775 K;
1 a.u. $\simeq$ 2.80 10$^{-17}$ cm$^2$; 1 a.u. $\simeq$ 6.126 10$^{-9}$ cm$^3$/s; 
1 a.u. $\simeq$ 5.1422 $10^{6}$ kV/cm.
}
\label{tab:systems}
\begin{tabular}{cccccccccccccccccccccc}
\hline\hline
                && $\mu$(a.u.)  && $B$(10$^{-7}$ a.u.) && $d$ (a.u.) && $s_{r_3}$($a_0$) && $s_{E_3}$(K) && ${E_{QR}}$(K) &&  $s_{\sigma_3}$(cm$^2$) && $s_{\beta_3}$(cm$^3$/s) && $\tilde{B}$ && $F$(kV/cm) \\ 
        &&  && && && && && \footnotesize{(at $\tilde{F}=3.4$)} && && && && \footnotesize{$\tilde{F}=[3.25 - 3.8]$} \\ \hline  
             $^1\Sigma$ &&&&&&&&&&&&&&&&&&&& \\ \hline  
 $^7$Li$^{23}$Na  && 27349  && 19.4 && 0.200 && 2188 && 1.2 10$^{-6}$ && 4.2 10$^{-3}$ && 1.68 10$^{-9}$ && 6.16 10$^{-9}$ && 5.07 10$^5$ && [161.8 -  189.2] \\ 
 $^{41}$K$^{87}$Rb  && 116547  && 1.67 && 0.226 && 11888 && 9.6 10$^{-9}$ && 3.4 10$^{-5}$ && 4.97 10$^{-8}$ && 7.85 10$^{-9}$ && 5.48 10$^6$ && [12.3 - 14.4] \\ \hline 
 $^{87}$Rb$^{133}$Cs  &&  200349 && 0.77 && 0.49 && 96207 && 8.5 10$^{-11}$ && 3.0 10$^{-7}$ && 3.26 10$^{-6}$ &&  3.70 10$^{-8}$ && 2.87 10$^8$ && [2.64 - 3.08] \\ 
 $^{23}$Na$^{41}$K  &&  58288 && 4.28 && 1.12 && 146234 && 1.3 10$^{-10}$ && 4.4 10$^{-7}$ && 7.52 10$^{-6}$ && 1.93 10$^{-7}$ && 1.07 10$^9$ && [6.39 - 7.47] \\ 
 $^{41}$K$^{133}$Cs  &&  158470 && 1.37 && 0.75 && 178279 && 3.1 10$^{-11}$ && 1.1 10$^{-7}$ && 1.12 10$^{-5}$ && 8.66 10$^{-8}$ && 1.38 10$^{9}$ && [3.05 - 3.56] \\ 
 $^7$Li$^{41}$K  && 43729  && 13.4 && 1.39 && 168978 && 1.3 10$^{-10}$ && 4.4 10$^{-7}$ && 1.01 10$^{-5}$ && 2.98 10$^{-7}$ && 3.33 10$^{9}$ && [16.05 - 18.77] \\ 
 $^{23}$Na$^{87}$Rb  && 100167 && 3.19  && 1.35 && 365108 && 1.2 10$^{-11}$ && 4.1 10$^{-8}$ && 4.69 10$^{-5}$ && 2.81 10$^{-7}$ && 8.52 10$^{9}$ && [3.95 - 4.62] \\ 
 $^7$Li$^{87}$Rb  && 85608  && 11.57 && 1.63 && 454902 && 8.9 10$^{-12}$ && 3.1 10$^{-8}$ && 7.28 10$^{-5}$ && 4.09 10$^{-7}$ && 4.10 10$^{10}$ && [11.87 - 13.87] \\ 
 $^{23}$Na$^{133}$Cs  && 142090  && 2.64 && 1.85 && 972605 && 1.2 10$^{-12}$ && 4.1 10$^{-9}$ && 3.33 10$^{-4}$ && 5.27 10$^{-7}$ && 7.10 10$^{10}$ && [2.39 - 2.79] \\ 
 $^7$Li$^{133}$Cs  && 127531  && 9.93 && 2.15 && 1179020 && 8.9 10$^{-13}$ && 3.1 10$^{-9}$ && 4.89 10$^{-4}$ &&  7.12 10$^{-7}$ && 3.52 10$^{11}$ && [7.72 - 9.03] \\ \hline 
             $^2\Sigma$ &&&&&&&&&&&&&&&&&&&& \\ \hline  
 $^{87}$Rb$^{84}$Sr  && 155695  && 0.82 && 0.606 && 114309 && 7.8 10$^{-11}$ && 2.7 10$^{-7}$ && 4.60 10$^{-6}$ && 5.65 10$^{-8}$ && 3.34 10$^8$ && [2.26 - 2.65] \\                
 $^{84}$Sr$^{19}$F  &&  93798 && 11.43 && 1.365 && 349535 && 1.4 10$^{-11}$ && 4.8 10$^{-8}$ && 4.30 10$^{-5}$ && 2.87 10$^{-7}$ && 2.62 10$^{10}$ && [13.99 -  16.36] \\                
 $^{138}$Ba$^{19}$F  && 143009  && 9.84 && 1.247 && 444881 && 5.6 10$^{-12}$ && 2.0 10$^{-8}$ && 6.97 10$^{-5}$ && 2.40 10$^{-7}$ && 5.57 10$^{10}$ && [13.18 -  15.41] \\                
 $^{89}$Y$^{16}$O  && 95611 && 17.68 && 1.78 && 605801 && 4.5 10$^{-12}$ && 1.6 10$^{-8}$ && 1.29 10$^{-4}$ && 4.88 10$^{-7}$ && 1.24 10$^{11}$ && [16.60 - 19.41] \\                
\hline\hline
\end{tabular}
\end{center}
\end{table*}

We consider ultracold identical bosonic molecules prepared initially in the state 
$|\tilde{n}_1 m_{n_1} \rangle| \tilde{n}_2 m_{n_2} \rangle$ =
$|\tilde{1}0\rangle|\tilde{1}0\rangle$. 
Another state $|\tilde{0}0\rangle|\tilde{2}0\rangle$ crosses the initial state
at a rescaled field $\tilde{F} = 3.25$. 
The energy curves of these states are indicated in the inset of Fig.~\ref{fig:nrg2}, 
where the initial energy ${E}_{\mathrm{init}}$ is indicated as a red bold dashed line. 
It has been shown and explained 
\cite{Avdeenkov_PRA_73_022707_2006,Wang_NJP_17_035015_2015,Quemener_PRA_93_012704_2016}
that the quenching processes were suppressed compared to the elastic ones, slightly 
beyond this field. We are then interested in the molecule-molecule scattering properties around this field. 
We assume the worst scenario for the molecules:
when the two molecules are suffciently 
close to each other they disappear from the experimental trap.
This can be due for example to a chemically reactive collision 
\cite{Zuchowski_PRA_81_060703_2010},
inelastic transitions to other states, 
collisional losses mediated by a long-lived complex 
\cite{Mayle_PRA_85_062712_2012, Mayle_PRA_87_012709_2013}.
In our calculations, this is satisfied when full loss $p_\mathrm{SR} = 1$
is invoked in Eq.~\eqref{eq:logder}.  
Thus the starting diagonal elements of the log-derivative matrix for a given channel 
are purely imaginary and given by $Z = - i \, k_\mathrm{min}$. 
We used $n_{1,2} = [0-3]$ for the rotational basis set.
We used $l = [0-10]$ for the partial wave basis employed in Eq.~\eqref{eq:kket}.
As we consider initial molecules in indistinguishable states, 
only symmetric states with $\eta = +1$ must be taken into acount in Eq.~\eqref{eq:kket}.
As we consider identical bosonic molecules, then $\varepsilon_P=+1$. 
The selection rules $\eta \, (-1)^l = \varepsilon_P$ implies even partial waves $l$.
The projection quantum number $M = m_{n_1} + m_{n_2} + m_l$ 
of the total angular momentum on the quantization axis
is conserved during the collision. 
We performed calculations for $M=0$ 
since the initial $m_{n_1} = m_{n_2} = 0$ 
and $m_l = 0$ is the dominant projection at ultralow collision energies.
In the following, we employ an arbitrary fixed rotational constant $B^* = 10^{-7}$ a.u.~($\sim 0.2$ cm$^{-1}$) and 
electric dipole moment $d^* =1$ a.u.~($\sim 2.54$ Debye) while the mass $\mu^*$ is varied 
in order to vary the parameter 
$\tilde{B} = 8 (B^*/\mathrm{a.u.}) (\mu^*/\mathrm{a.u.})^3 (d^*/\mathrm{a.u.})^4$
in Eq.~\eqref{eq:Btilde}.
The star characterizes a hypothetical dipolar molecule, say XY$^*$, 
defined by those values which also define 
a characteristic length $s_{r_3^*}$, energy $s_{E_3^*}$, 
cross section $s_{\sigma_3^*}$, and
 rate coefficient $s_{\beta_3^*}$.
Fixing $B^*$ and $d^*$ is also convenient
for varying the rescaled field 
$\tilde{F} = d^*F/B^*$, since 
it is sufficient to vary the electric field $F$
only.
We consider the scattering properties at collision energies ${E}_{c}^* = 100$~nK
so that the third parameter $\tilde{E}_{c}$ is fixed.
We used $r_\mathrm{min} = 5 \, a_0$ and $r_\mathrm{max}$ 
is chosen so that ${{k}}^* \, r_\mathrm{max}^* \sim 5$.
As the mass $\mu^*$ is changed here to vary the parameter $\tilde{B}$, ${{k}}^*$ changes 
accordingly, and so does $r_\mathrm{max}^*$.
Most of the systems investigated 
in experiments are diatomic dipolar molecules of alkali atoms
for which the electronic $C_6^{\mathrm{el}}$ coefficients belongs 
to the range $-20000 \le C_6^{\mathrm{el}} \le -3000$ a.u.
\cite{Kotochigova_NJP_12_073041_2010,Lepers_PRA_88_032709_2013,Vexiau_JCP_142_214303_2015,Kotochigova_NJP_12_073041_2010,
Zuchowski_PRA_87_022706_2013}. 
In this study we use a fixed value of $C_6^{\mathrm{el},*} = -10000$ a.u.~between
two molecules XY$^*$.
We discuss the effect of the $C_6^{\mathrm{el}}$ coefficient at the end of this section.
We obtain the rescaled scattering length $\tilde{a}$ and all related quantities (see Eq.~\eqref{eq:rescaledscat}) by dividing the scattering length $a^*$ with $s_{r_3^*}$ computed 
for the hypothetical molecule XY$^*$, 
for different values of $\mu^*$ and $F^*$ correponding to different values
of $\tilde{B}$ and $\tilde{F}$.

\begin{figure*}[t]
\includegraphics[width=86mm]{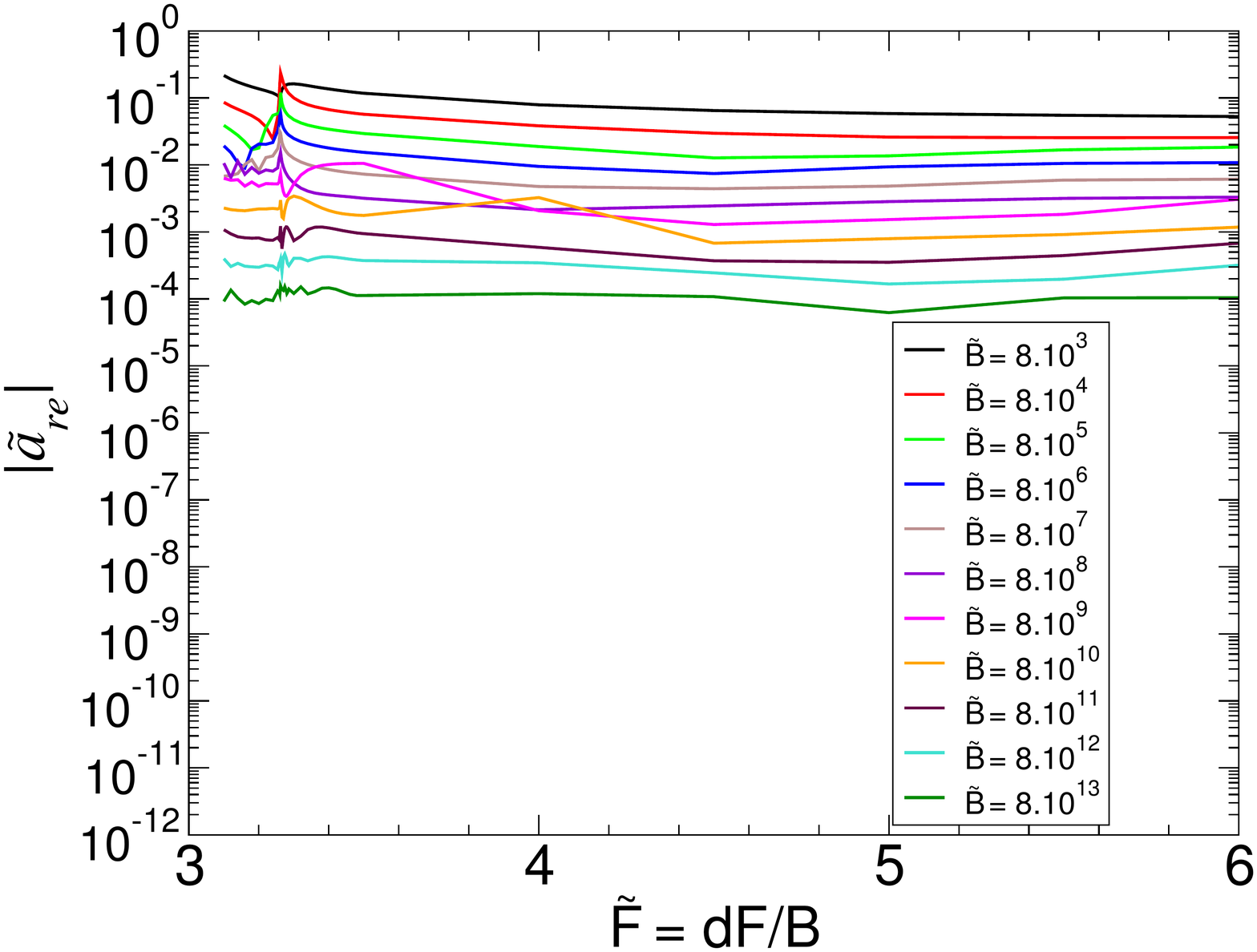}
\includegraphics[width=86mm]{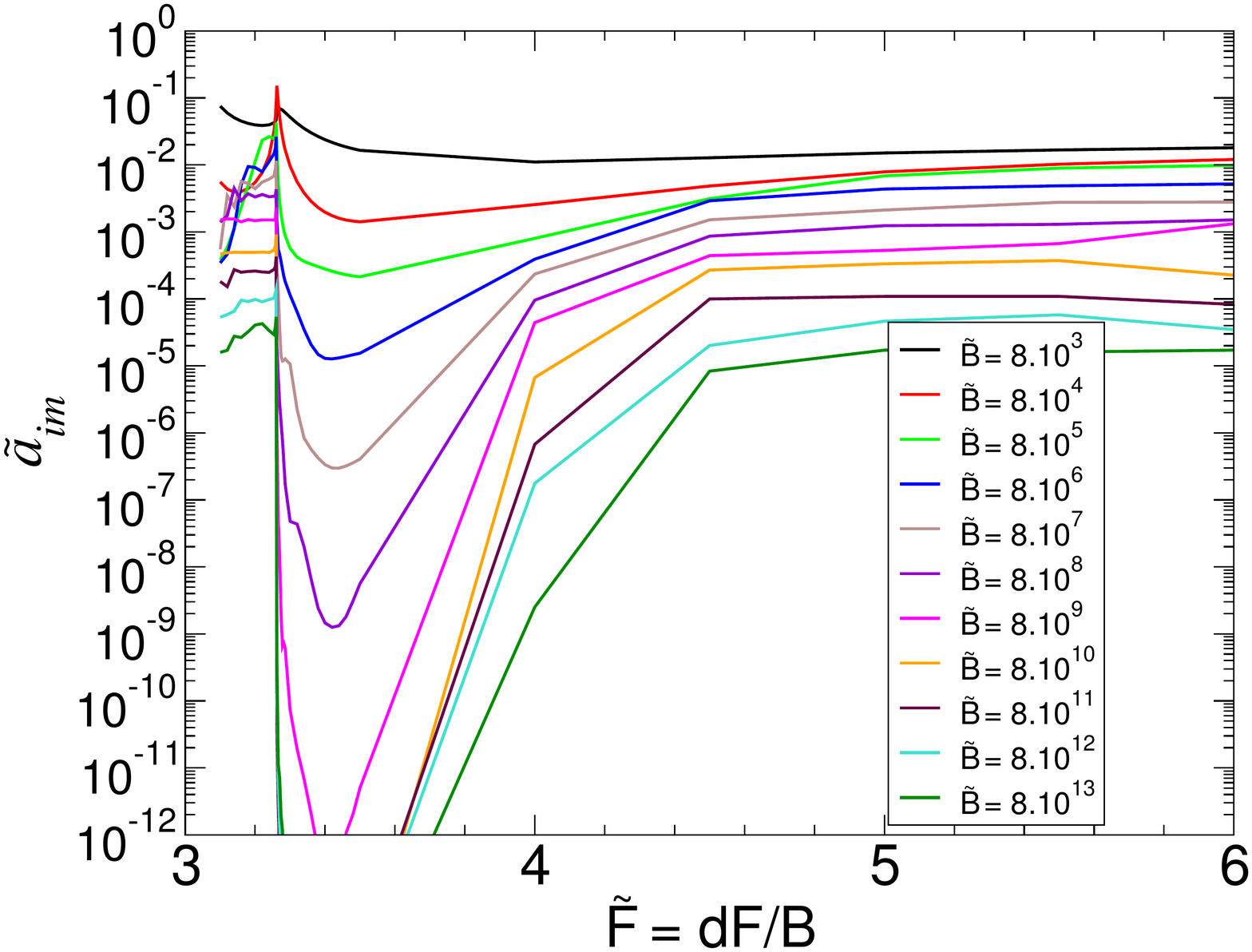}
 \caption{(Color online). Absolute value of the real part $|\tilde{a}_\mathrm{re}|$
 (left panel) and imaginary part $\tilde{a}_\mathrm{im}$ (right panel) 
 of the rescaled scattering length
 as a function of $\tilde{F}$ for different values of $\tilde{B}$.}
 \label{fig:are-aim}
\end{figure*}

The quantity $|\tilde{a}|^2/\tilde{a}_\mathrm{im} \equiv \gamma/\tilde{k}$ 
is plotted in Fig.~\ref{fig:asqoveraim} 
as a function of $\tilde{B}$ and $\tilde{F}$. 
Different contour plots are drawn 
from dark blue for low values of this quantity ($10^{-4}$) to dark red
for high values ($10^{6}$). White contour plots correspond to value $\ge 10^{6}$.
When multiplying $|\tilde{a}|^2/\tilde{a}_\mathrm{im}$
by the rescaled wavevector ${\tilde{k}}$, 
this provides the ratio $\gamma$ 
for the collision energy $\tilde{E}_c$, see Eq.\eqref{eq:rescaledscat}. 
Therefore, this plot gives directly the ratio $\gamma$
for ${\tilde{k}}=1$ that is when $E_c = s_{E_3}$. 
For efficient evaporative cooling to occur, 
a ratio $\gamma \ge 100$ \cite{Cornell_RMP_74_875_2002}, 
and perhaps a safer value of $\gamma \ge 1000$, is required.
The latter condition corresponds in the figure to the 
orange, red, dark red, and white contour plots.
Therefore, the condition
for favorable evaporative cooling
is delimited approximately by the region $\tilde{B} \ge 10^8$
and $3.25 \le \tilde{F} \le 3.8$, 
with a maximum ratio reached for $\tilde{F} \simeq 3.4$ for a given $\tilde{B}$.
Any other position in the plot is likely to be
unfavorable.
This universal feature is due to the shielding mechanism
\cite{Bohn_PRA_63_052714_2001,Avdeenkov_PRA_73_022707_2006,Wang_NJP_17_035015_2015,
Quemener_PRA_93_012704_2016}
when the incident collisional channel becomes repulsive enough so that the quenching rate coefficient is supressed.

The characteristic values of the dipolar bi-alkali molecules
are reported in Tab.~\ref{tab:systems}. The $\tilde{B}$ values
are reported on the right of Fig.~\ref{fig:asqoveraim} as yellow dashed lines.
This distinguishes two groups of molecules 
for evaporative cooling, 
the good candidates from the bad.
Group 1 (RbCs, NaK, KCs, LiK, NaRb, LiRb, NaCs, LiCs)
for which $\tilde{B} > 10^8$ has favorable candidates, while
group 2 (LiNa, KRb) has unfavorable ones since $\tilde{B} \ll 10^8$.
This holds at collision energies $E_c = s_{E_3}$ 
(see Tab.~\ref{tab:systems} for the values),
when ${\tilde{k}}=1$. 
As mentionned in Sec.~\ref{sec:subsecD}, a more appropriate value
is when the collision energy is on the order of 
the quantum regime energy ($E_c = E_{QR}$) since 
it better reflects the magnitude of the interaction and the collision
for the given applied field. 
Let's take the example of $\tilde{F} = 3.4$.
At this field, $d_\mathrm{ind} \simeq 0.13 \ d$ (this can be directly
calculated from Fig.~\ref{fig:nrg1} using the slope of the 
$|\tilde{1}0\rangle$ curve at $\tilde{F} = 3.4$).
Then $E_\mathrm{QR}(\tilde{F}=3.4) \simeq s_{E_3}/0.13^{4} \simeq 3500 \, s_{E_3}$.
The corresponding values for each molecule are reported in Tab.~\ref{tab:systems}.
If now $E_c = E_{QR}$, the ratio should become
$\gamma(\tilde{F}=3.4) = \tilde{k}_{QR} \times |\tilde{a}|^2/\tilde{a}_\mathrm{im}$
with $\tilde{k}_{QR} = \sqrt{E_{QR}/s_{E_3}} \simeq \sqrt{3500} \simeq 60$.
The ratio should increase by a factor of 60 for this example
compared to the one for $E_c = s_{E_3}$.
The precedent conclusions remain unchanged
since for the first group, the ratio $\gamma$ will be bigger than 1000
while for the second group, the ratio increases by the factor of 60 but
is not enough to reach the ratio of $1000$.

The white contour plots in Fig.~\ref{fig:asqoveraim}
correspond to values of the ratio bigger than $10^6$
at ${\tilde{k}}=1$. This area is not shown in more detail 
since we encounter numerical 
issues leading to unphysical oscillations in the values of the scattering quantities.
In this region, the quenching processes are so 
strongly suppressed that 
the values of $\tilde{a}_\mathrm{im}$ compared to the ones of 
$|\tilde{a}_\mathrm{re}|$ are very tiny, 
about $10^{-10}$ smaller (see Fig.~\ref{fig:are-aim} below). 
We believe the log-derivative method cannot achieve 
higher precision
and produces numerical errors.
One can use for example more appropriate methods for better numerical precision
\cite{Simoni_arXiv_1705_4102_2017} to fulfill the plot in the white area. 
From an experimental point of view though, the ratio presented 
in the figure is already more than sufficient.
When $\tilde{a}_\mathrm{im} \ll \tilde{a}_\mathrm{re}$, $|\tilde{a}|^2 \simeq |\tilde{a}_\mathrm{re}|^2$
so that $|\tilde{a}|^2/\tilde{a}_\mathrm{im} \simeq |\tilde{a}_\mathrm{re}|^2/\tilde{a}_\mathrm{im}$.
Since $|{a}_{re}|/\tilde{a}_\mathrm{im} \sim 10^{10}$ and $|{a}_{re}| \ge 10^{-4}$ 
(see Fig.~\ref{fig:are-aim}), then the white area corresponds to
$|\tilde{a}_\mathrm{re}|^2/\tilde{a}_\mathrm{im} \ge 10^{6}$.

The results in Fig.~\ref{fig:asqoveraim} are promising 
for bosonic dipolar molecules under current experimental interest,
such as $^{87}$Rb$^{133}$Cs \cite{Takekoshi_PRL_113_205301_2014,Molony_PRL_113_255301_2014} and $^{23}$Na$^{87}$Rb
\cite{Guo_PRL_116_205303_2016} since they belong to the first group as defined above.
For NaRb, at $\tilde{F} \simeq 3.4$, 
the ratio $\gamma$ reach values above 10$^6$ 
for the collision energy range from $s_{E_3} = 1.2 \ 10^{-11}$~K 
($\tilde{k}=1$) to $E_{QR} = 4.1 \ 10^{-8}$~K. 
This is then well appropriate 
to reach quantum degeneracy of ultracold dipolar gases 
and form Bose-Einstein condensates of dipolar molecules. 
To compare with, the typical critical temperature
$T_c \sim 3.3125 \, \hbar^2 \, n^{2/3} / m k_B$ 
where condensation takes place (though for a ideal non-interacting Bose gas) 
with a typical density of $10^{12}$ molecules/cm$^3$ is $T_c \sim 10$~nK
for NaRb, which belongs to the energy range where the ratio is favorable for evaporative cooling.
It should be noted that $E_{QR}$ is not an upper limit of
the collision energy above which the evaporative cooling technique
would become unfavorable. 
The ratio can still remain big (above 1000) for even higher collision energies.
It just means that one cannot strictly use Eq.\eqref{eq:ratio} to convert
the quantity $|\tilde{a}|^2/\tilde{a}_\mathrm{im}$ to the ratio $\gamma$
using $\tilde{k}$. 
For instance for NaRb, the ratio is above 10$^6$ at $E_c = E_{QR}$,  but 
it can still take a high collision energy for the ratio to get down to 1000. 
To answer up to which collision energy for each system, 
one has then to repeat the calculation presented in Fig.~\ref{fig:asqoveraim} 
as a function of the collision energies. This is not shown here
but can be calculated upon request (and for a specific system to save computational time).
For RbCs, the ratio $\gamma$ can reach values of 1000 or above
for the collision energy range from $s_{E_3} = 8.5 \ 10^{-11}$~K  
to $E_{QR} = 3 \ 10^{-7}$~K, but at a somewhat more limited 
range of electric fields as shown in Fig.~\ref{fig:asqoveraim},
around $\tilde{F} \simeq 3.4$. Above or below this field, the ratio can decrease
and could become unfavorable.

The results of this paper are not necessarily constrained to bosonic molecules. 
Earlier studies \cite{Avdeenkov_PRA_73_022707_2006,Wang_NJP_17_035015_2015} 
showed that fermionic molecules also experience quenching suppression.
In addition, the adimensional study and parameters remain valid,
so that similar outcomes are expected for identical fermionic molecules. 
Examples of fermionic dipolar molecules of current experimental interest
are $^{40}$K$^{87}$Rb \cite{Ni_S_322_231_2008} 
and $^{23}$Na$^{40}$K \cite{Park_PRL_114_205302_2015}. 
While fermionic KRb are not good candidates (this was shown already in Ref. \cite{Wang_NJP_17_035015_2015}), we can expect that NaK will be a good candidate for quenching suppression.
In contrast with fermionic neutral particles (alkali atoms, homonuclear molecules)
which interact via the van der Waals interaction at long-range, 
fermionic dipolar particles in an electric field interact via the dipolar interaction. 
This modifies the $p$-wave threshold laws of the elastic process 
\cite{Sadeghpour_JPBAMOP_33_93_2000,Bohn_NJP_11_055039_2009}. 
The elastic cross section tends to a constant at vanishing collision energies,
in contrast with the van der Waals interaction where the elastic cross section
vanishes as ${E_c}^2$. Since the quenching 
cross section behaves as $\sqrt{E_c}$, the ratio $\gamma$ increases 
at ever lower collision energies. 
Therefore successful evaporative cooling can also be used to reach quantum degeneracy
and form degenerate Fermi gases of dipolar molecules.

Another important experimental issue is the range of fields 
at which the suppression takes place, reported as $F_{\tilde{F}=[3.25 - 3.8]}$ in the last column
of Tab.~\ref{tab:systems}. 
For example the LiNa system would require too high electric fields,
above 100 kV/cm, to implement the already weak suppression. 
Generally in an experiment, electric fields 
up to $\sim$ 5 kV/cm can be created when the electrodes stand outside the vacuum chamber
\cite{Ni_N_464_1324_2010}. Therefore, the suppression can be implemented in such circumstances
for the RbCs, KCs, NaRb, NaCs systems, which require electric fields smaller than 5 kV/cm.
For the remaining systems KRb, NaK, LiK, LiRb, LiCs, higher fields are required, and
the electrodes must be included inside the vacuum chamber \cite{Moses_NP_13_13_2017}.

The characteristic values of representative $^2\Sigma$ dipolar molecules 
such as RbSr, SrF, BaF, YO, which are also of experimental interest \cite{Barry_N_512_286_2014,McCarron_NJP_17_035014_2015,Pasquiou_PRA_88_023601_2013,Hummon_PRL_110_143001_2013, Collopy_NJP_17_055008_2015,Yeo_PRL_114_223003_2015},
are reported in Tab.~\ref{tab:systems}.
These molecules are not perfect Hund's case b type of molecules since they have an additional fine and
hyperfine structure that should be included in the Hamiltonian.
Nevertheless, around the electric field $\tilde{F}=3.25$, the electronic and nuclear spins 
can mainly act as spectators \cite{Quemener_PRA_93_012704_2016}.
The corresponding $\tilde{B}$ values, 
reported as red lines on the left of Fig.~\ref{fig:asqoveraim},
show that for SrF, BaF and YO, the quantity $|\tilde{a}|^2/\tilde{a}_\mathrm{im}$ is well above $10^3$,
making them potential candidates for successful evaporative cooling under the asumption 
that the spins are spectators. 
For RbSr, this is like RbCs as discussed above since they share a similar value of $\tilde{B}$.
$\gamma$ can reach 1000 but at a restricted range of electric fields.

One cannot tell from Fig.~\ref{fig:asqoveraim} whether high ratios are due
to high values of $\tilde{a}$, low values of $\tilde{a}_\mathrm{im}$, or a combination of both.
This can be seen in Fig.~\ref{fig:are-aim} which shows 
the absolute value of the real part $|\tilde{a}_\mathrm{re}|$ (left)
and imaginary part $\tilde{a}_\mathrm{im}$ (right) of the rescaled 
scattering length as a function of $\tilde{F}$, for different values of $\tilde{B}$,
from $8. \, 10^{3}$ to $8. \, 10^{13}$.
$|\tilde{a}_\mathrm{re}|$ does not vary much with $\tilde{F}$ while $\tilde{a}_\mathrm{im}$ does.
When the field crosses the value $\tilde{F}^* = 3.25$ it strongly suppresses the quenching 
processes while the elastic ones remain relatively steady. 
The reason for the high ratio comes then from a suppressed value of $\tilde{a}_\mathrm{im}$
rather than an enhanced value of $|\tilde{a}_\mathrm{re}|$.
These two plots are also useful to have a direct magnitude
of the quenching rate coefficients and elastic cross sections.
$\tilde{a}_\mathrm{im}$ gives the quenching rate coefficients when multiplied by 
$s_{\beta_3} \times \Delta$ (see the values in Tab.~\ref{tab:systems})
while $|\tilde{a}_\mathrm{re}|^2$, when $\tilde{a}_\mathrm{im} \ll \tilde{a}_\mathrm{re}$, 
does the same for the elastic cross sections 
when multiplied by $s_{\sigma_3} \times \Delta$ (see Eq.~\eqref{eq:rescaledscat}).

\begin{figure}
 \includegraphics[width=86mm]{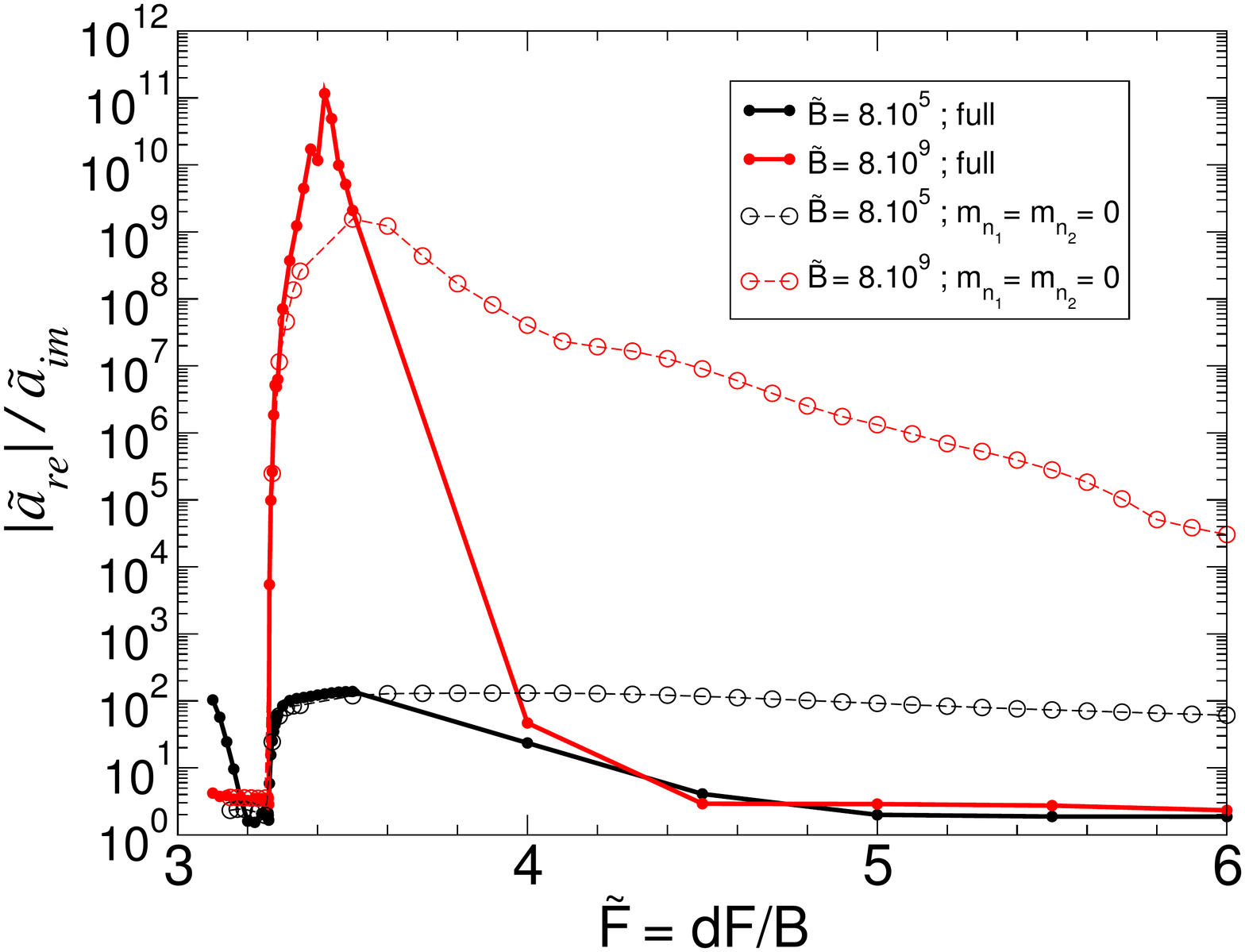}
 \caption{(Color online).
 Ratio $|\tilde{a}_\mathrm{re}|/\tilde{a}_\mathrm{im}$
as a function of $\tilde{F}$ for $\tilde{B} = 8. \, 10^{5}$ (black curves) 
and $\tilde{B} = 8. \, 10^{9}$ (red curves).
Solid curves: full calculation; dashed curves: $m_{n_1} = m_{n_2} = 0$ approximation.}
 \label{fig:M0test}
\end{figure}

Fig.~\ref{fig:M0test} confirms a useful information on the mechanism
of the quenching suppression. As mentioned in \cite{Quemener_PRA_93_012704_2016},
a useful approximation nearby $\tilde{F}=3.25$ consists in taking only the $m_n = 0$ projection
of the molecules in the calculation ($m_{n_1} = m_{n_2} = 0$). 
This is verified from this figure, which shows 
the ratio $|\tilde{a}_\mathrm{re}|/\tilde{a}_\mathrm{im}$
as a function of $\tilde{F}$ for two values of 
$\tilde{B} = 8. \, 10^{5}$ (black) and $8. \, 10^{9}$ (red).
The solid curves result from a full calculation, as also shown in the previous figures,
while the dashed curves result from the approximation $m_{n_1} = m_{n_2} = 0$ 
for the rotational states of the molecules.
As one can see in Fig.~\ref{fig:M0test}, the approximation is valid 
for fields in the range $3.25 \le \tilde{F} \le 3.3$.
For higher fields, the approximation becomes less and less valid. 
It strongly overestimates the results at larger fields.
The approximated calculation is much faster than the full calculation as it takes much less 
molecular states into account in the scattering, decreasing the size of the coupled equations.
The corresponding molecular states are indicated as red lines in Fig.~\ref{fig:nrg2}, while
the full calculation employs all the curves (red and black).
Therefore this approximation is worth using  
at fields in the range $3.25 \le \tilde{F} \le 3.3$
especially due to its numerical simplicity.
This range is somewhat restricted in field but even at $\tilde{F} = 3.3$ 
it can indicate with not much numerical effort that the suppression can be already quite strong.

\subsection*{Effect of the electronic van der Waals coefficient}

\begin{figure}
 \includegraphics[width=86mm]{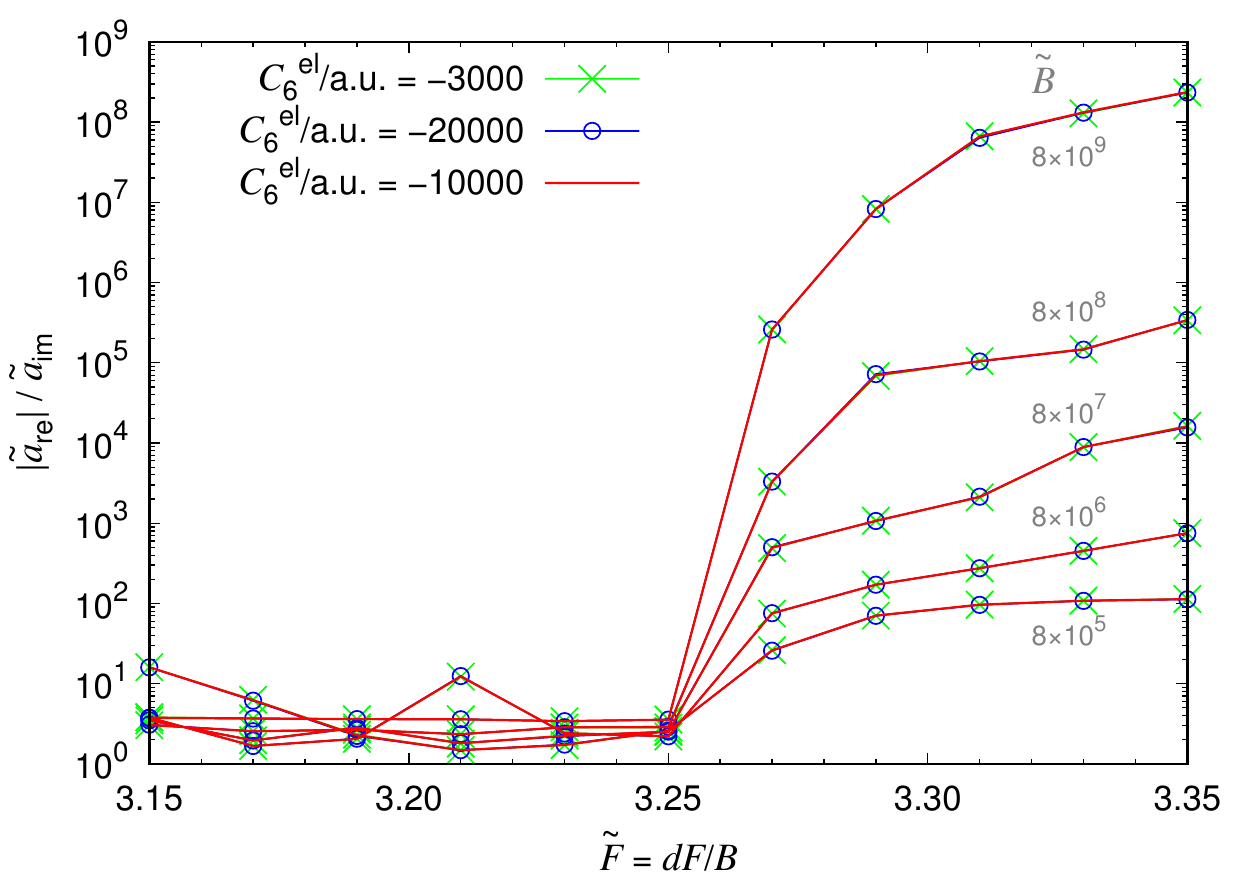}
 \caption{(Color online).
 Ratio $|\tilde{a}_\mathrm{re}|/\tilde{a}_\mathrm{im}$
as a function of $\tilde{F}$ for different values of $\tilde{B}$.
Solid red curves: $C_6^\mathrm{el,*}=-10000$ a.u.~, green cross curves: 
$C_6^\mathrm{el,*}=-3000$ a.u.~, blue circle curves: $C_6^\mathrm{el,*}=-20000$ a.u.}
 \label{fig:C6test}
\end{figure}

\begin{table}[!h]
\setlength{\extrarowheight}{7pt}
 \begin{center}
  \caption{Van der Waals $C_6$ coefficients for different systems. 
  $C_6^{\mathrm{rot}} \simeq (d/\mathrm{a.u.})^4/(B/\mathrm{a.u.})$ 
  is the repulsive ``rotational"  van der Waals coefficient
  responsible for the repulsive interaction \cite{Wang_NJP_17_035015_2015}. 
  $C_6^{\mathrm{el}}$ is the ``electronic" van der Waals coefficient, taken from   
  \cite{Lepers_PRA_88_032709_2013}.
  The two last columns are the rescaling factor and the rescaled 
  electronic van der Waals coefficient
  $C_6^{\mathrm{el,resc.}}$ from the fixed coefficient $C_6^{\mathrm{el},*} = -10000$ a.u. 
  used in our study (see text for details). 
1 a.u. of $C_6$ = 1 $E_h.a_0^6$ where $E_h$ is a Hartree and $a_0$ the Bohr radius.
  \label{tab:C6}}
  \begin{tabular}{cccccccccc}
   \hline\hline
&& $C_6^{\mathrm{rot}}$ (a.u.)  && $C_6^{\mathrm{el}}$ (a.u.) && $\frac{s_{E_3} s_{r_3}^6}{s_{E^*_3} s_{r^*_3}^6}$  && $C_6^{\mathrm{el,resc.}}$ (a.u.)  \\ \hline  
 $^7$Li$^{23}$Na  && 826 && -3342 && 0.00008 && -0.8   \\ 
 $^{41}$K$^{87}$Rb && 15623 && -12636 && 0.0016 && -16   \\ \hline 
 $^{87}$Rb$^{133}$Cs && 744251 && -17760 &&  0.074 && -744 \\ 
 $^{23}$Na$^{41}$K  && 3673910 && -7532 &&  0.37 && -3674  \\ 
 $^{41}$K$^{133}$Cs  &&  2314772 && -16230 &&  0.23  && -2315 \\ 
 $^7$Li$^{41}$K  && 2796249 && -6689 && 0.28 && -2796    \\ 
 $^{23}$Na$^{87}$Rb  && 10414091 && -9046 && 1.04 && -10414   \\ 
 $^7$Li$^{87}$Rb  && 6099595 && -8114 && 0.61 && -6100   \\ 
 $^{23}$Na$^{133}$Cs && 44324439 && -11998 && 4.43 && -44324  \\ 
 $^7$Li$^{133}$Cs  && 21512044 && -11007 && 2.15 && -21512   \\ 
   \hline\hline
  \end{tabular}
 \end{center}
\end{table}

Finally, we discuss the effect of the $C_6^{\mathrm{el}}$ coefficient. As mentioned previously 
in Sec.~\ref{sec:subsecC}, the study is in general not strictly adimensional because of 
the van der Waals $C_6^{\mathrm{el}}/r^6$ interaction term. 
But to which extent this is true?
This is what Fig.~\ref{fig:C6test} answers, where the ratio
$|\tilde{a}_\mathrm{re}|/\tilde{a}_\mathrm{im}$
is plotted as a function of $\tilde{F}$
for different values of $\tilde{B}$ and different values of $C_6^\mathrm{el,*}$.
The ratio does not change for the different $C_6^\mathrm{el,*}$ employed.
The reason can be understood as follows.
There are two competing effects for the dispersion term between two XY molecules: 
(i) an attractive ``electronic'' van der Waals interaction with a negative coefficient $C_6^{\mathrm{el}}$;
(ii) a ``rotational'' van der Waals interaction with a 
coefficient $C_6^{\mathrm{rot}}$ that can be tuned positive or negative depending on the electric field \cite{Wang_NJP_17_035015_2015}.
The former coefficients are taken from \cite{Lepers_PRA_88_032709_2013} between
two $|\tilde{0}0\rangle$ molecules and are negative since the interaction is attractive.
These coefficients constitute an upper value in magnitude
for the coefficient between two $|\tilde{1}0\rangle$ molecules.
The latter coefficient can be estimated semi-quantitatively 
by second order perturbation theory where the correction behaves as
$\simeq W^2/\Delta E$. The dipolar interaction scales 
as $W \simeq (d^2/4\pi\varepsilon_0)/r^3$. 
An upper value of the difference in energy between the states
$|\tilde{1}0\rangle|\tilde{1}0\rangle$ and $|\tilde{0}0\rangle|\tilde{2}0\rangle$ 
is approximately $\Delta E \simeq B$ for $\tilde{F} \ge 3.25$ (see Fig.~\ref{fig:nrg2}).
This provides an order of magnitude of the 
repulsive van der Waals interaction $ \simeq (d^2/4\pi\varepsilon_0)^2/(B r^6)$
with a positive $C_6^{\mathrm{rot}} \simeq (d/\mathrm{a.u.})^4/(B/\mathrm{a.u.})$ for the initial 
state $|\tilde{1}0\rangle|\tilde{1}0\rangle$.
Both values are reported in Tab.~\ref{tab:C6}.
The value we use is actually not $C_6^{\mathrm{el}}$  but 
$C_6^{\mathrm{el},*} = -10000$ a.u. as mentioned above for the hypothetical system XY$^*$. This is a fixed value.
However, to obtain the rescaled $C_6^{\mathrm{el,resc.}}$ coefficient 
for the real bi-alkali dipolar molecules,
we have to rescale $C_6^{\mathrm{el},*}$ with a rescaling factor so that
\begin{eqnarray}
C_6^{\mathrm{el,resc.}} = C_6^{\mathrm{el},*} \times \frac{s_{E_3} s_{r_3}^6}{s_{E^*_3} s_{r^*_3}^6} ,
\end{eqnarray}
which depends on the system. This is due to the fact 
that we use a characteristic length and energy 
relative to the dipolar interaction instead of the van der Waals interaction. 
The values of the rescaling factor and the resulting 
$C_6^{\mathrm{el,resc.}}$ from $C_6^{\mathrm{el},*} = -10000$ a.u.~are reported in Tab.~\ref{tab:C6}.
One can see that the $C_6^{\mathrm{el,resc.}}$ coefficients
are always much smaller than the $C_6^{\mathrm{rot}}$ ones
so it will not affect the scattering quantities, as seen in Fig.~\ref{fig:C6test}.
In that sense, the study can be considered as independent of this
coefficient and then adimensional for this specific $C_6^{\mathrm{el},*}$.

But is it appropriate to use the value $C_6^{\mathrm{el},*} = -10000$ 
a.u.~to describe the real molecules XY? 
And does it quantitatively affect the results? 
For the systems of group 1, 
although the values of $|C_6^{\mathrm{el,resc.}}|$ 
do not reproduce exactly the values of the real $|C_6^{\mathrm{el}}|$, 
this is still acceptable 
since they are much smaller than $|C_6^{\mathrm{rot}}|$.
In other words, 
as far as $|C_6^{\mathrm{el}}|$ remains small
compared to $|C_6^{\mathrm{rot}}|$,
the value of the scattering quantities are not going to be affected
with this value of $C_6^{\mathrm{el},*}$ employed, 
and the study is considered adimensional for this group.
This is questionable for group 2 though, where 
$|C_6^{\mathrm{el}}|$ 
is comparable or bigger than $|C_6^{\mathrm{rot}}|$, and one has to be careful with the value of $C_6^{\mathrm{el},*}$ used.
One can see that the $|C_6^{\mathrm{el,resc.}}|$ coefficients are much smaller than the real ones $|C_6^{\mathrm{el}}|$ so that we strongly underestimate their values in our calculation.
In contrast with group 1, 
this is not acceptable since we cannot neglect the value of 
$|C_6^{\mathrm{el}}|$ compared to the 
value of $|C_6^{\mathrm{rot}}|$.
Therefore, the scattering quantities 
and the ratio $\gamma$
are certainly affected and the study cannot be considered as adimensional
for group 2. A systematic study is then recommended
including the proper $C_6^{\mathrm{el}}$ coefficient.
But when doing so, 
for KRb for instance \cite{Wang_NJP_17_035015_2015},
the order of magnitude of the ratio $\gamma$ still remains far below 1000.
Then the definition of group 1 and 2 determined above remains unchanged.

\section{Conclusion}
\label{sec:conclusion}

In conclusion, we performed a general study 
on shielding ultracold dipolar rotors using an adimensional perspective,
in order to identify which systems are good candidates for efficient evaporative cooling
based on two-body collisions.
We showed that, among the bi-alkali dipolar molecules, two groups can be distinguished.
Group 1, including the molecules RbCs, NaK, KCs, LiK, NaRb, LiRb, NaCs, LiCs,
is favorable for efficient evaporative cooling using the shielding mechanism
as they have a ratio elastic/quenching processes over 1000 at 
a collision energy equal to and even higher than their characteristic dipolar energy.
Group 2, including LiNa and KRb, is not favorable.
In general, the study is not strictly adimensional since it contains two competing interactions, the electronic van der Waals interaction and the dipolar interaction,
from which different characteristic length and energy can be defined.
As we rescale the Schr{\"o}dinger equation with the dipolar length and energy,
the rescaled expression of the electronic van der Waals interaction breaks the adimensionality.
However, the study can be considered adimensional for the first group 
since the electronic van der Waals coefficient is small in magnitude 
compared to the rotational one, 
responsible for the shielding.
For group 2, it can not be considered adimensional as
the electronic van der Waals coefficient is comparable or even 
bigger in magnitude compared to the rotational one,
so that the electronic van der Waals coefficient we used
is underestimated. A systematic study is then recommended for group 2.
Despite that, the conclusions of the paper remained qualitatively
unchanged.

\begin{acknowledgments}
The authors acknowledge insightful discussions with the members of 
the Th{\'e}omol team at Laboratoire Aim{\'e} Cotton. 
M.L.G.-M. and G.Q. acknowledge funding from the COPOMOL 
project no.~ANR-13-IS04-0004 from Agence Nationale de la Recherche.
J.L.B. acknowledges funding from the JILA Physics Frontier 
Center and an ARO MURI Grant No. W911NF-12-1-0476.
\end{acknowledgments}

\bibliography{../../../BIBLIOGRAPHY/bibliography.bib}

\end{document}